# Software Engineering as Instrumentation for the Long Tail of Scientific Software

Daisie Huang, daisieh@mail.ubc.ca, Beaty Biodiversity Centre, University of British Columbia
Hilmar Lapp, hlapp@nescent.org, National Evolutionary Synthesis Center (NESCent)

## Introduction

Many disciplines of science are increasingly reliant on computational tools to extract insight from data. Some of these tools have the character of domain-agnostic infrastructure with potential use, and thus possibilities for continued support, from a wide swath of research fields, or even disciplines. Nonetheless, any quantitative research field will also have a myriad of algorithms for analysis and visualization of its data, all implemented in different software tools. For the field of evolutionary biology alone, a website [1] maintained by one of its most renowned scientists lists more than 350 analysis and visualization programs. The vast majority of this long tail of scientific software is highly specialized and purpose-built for a research project. Once the research project concludes, further sustaining the development and maintenance of its software products typically has to rely on community uptake and reuse. When opportunities for reuse arise or whether uptake will take place at all is often difficult to predict, but a variety of the factors that influence a potential new user's or new developer's decision on whether to adopt an existing tool or start a new one are well known. Many of these factors relate to the ease, or difficulty, with which a new user can understand a tool, with which it can be used in situations it was not originally designed for, or with which new capabilities can be added to it. Removing these kinds of difficulties is the subject of the science of software engineering. For a variety of reasons employing software engineering practices as well as talent as part of the development of long tail scientific software has proven to be challenging. As a consequence, community uptake of long tail tools is often far more difficult than it would need to be. At the same time, tools that recreate the functionality of another while adding some incremental improvements are a frequent phenomenon in science, signifying that opportunities for reuse abound.

Exploring different approaches to the same problem is common in science, and having multiple specialized tools with largely the same functionality could be seen as a part of this valued tradition. However, a new algorithm, let alone a better implementation of an existing one, does not necessarily require a new tool to be written, and many of the tasks a tool has to perform to be functional are not unique to a particular algorithm. In the absence of community adoption, tools in the long tail face the risk of being lost, because due to finite financial resources funding agencies cannot provide continued support beyond the initial innovation phase. For some tools their purpose may be narrow enough that their loss may not be felt widely. Even so, the long tail of software tools as a whole is what research fields critically depend on as the place for computational innovation that enables new insights and new research.

Here, we briefly discuss likely reasons why employing software engineering in the long tail is challenging, and propose that many of those obstacles could be addressed in the form of a cross-cutting non-profit center of excellence that makes software engineering broadly accessible as a shared service, conceptually and in its effect similar to shared instrumentation.

## Why engineering for reuse is necessary yet challenging for the long tail

Despite the impressive success of some open-source projects in sustaining vibrant developer and user communities for more than a decade (e.g., [2,3]), it is increasingly recognized that an open-source license may be necessary, but is far from sufficient for fostering community uptake. Reuse of a tool by scientists in the form of applying it to their own data or question may be hampered by poor user interfaces, lack of scalability, or difficulty to interoperate with other tools. A potential new developer may be dissuaded from adding new or improving existing capabilities if integrity tests are missing, or if the design of the code and information flow is hard to understand. That such barriers to reuse are common for long tail tools is a result of most of them having been designed for a narrow purpose and not for wide reuse. Furthermore, long tail tools are by and large developed by researcher-developers whose career priorities, scientific interests, and training are primarily in their domain science. Successfully optimizing software to reduce or eliminate barriers to reuse takes training in the science of software engineering and its subfields such as developer productivity, infrastructure engineering, user testing, user experience, and performance engineering, which even more experienced researcher-developers often do not have. Those that do are not rewarded for it by the academic incentive structure, which by and large provides little recognition for effort devoted to engineering improvements unless it is immediately evident how such effort enables new science. Correspondingly, among many domain scientists, awareness of software engineering benefits and appreciation for its cost are low. Investigators applying for grants often do not include funding for project-relevant software engineering effort, and if they do, review panels often question the necessity of such funds [4]. In this climate, the human talent in software engineering that academia has is eventually lost to industry, often at the peak of their training, in part because universities offer little in the way of alternative career tracks for such people.

There are other factors lowering (or creating) barriers to uptake, in particular nurturing community through outreach, engagement, and responsiveness. Addressing these well is hampered by similar issues as the software engineering-related ones: it takes effort by human resources with rarefied skills, and it is typically not recognized by the academic incentive system.

**A Possible Solution: A non-profit software engineering center of excellence**

In many ways, infusing software engineering resources into scientific software development is an investment for future returns in the form of community uptake and reuse. For infrastructure-level projects these long term returns may be well understood and part of the project mission to begin with. However, for tools in the long tail this investment will be more difficult to justify, especially if, as illustrated above, the costs of investing are high, and the anticipated returns may not manifest in the end. Although there could be immediate benefits of better engineering, journals, science funders, and even scientists themselves typically do not reward, let alone mandate them. Shifting the balance from costs to benefits thus takes two kinds of changes. One is changing the reward structure to better recognize the immediate and near-term benefits of well-engineered software. The other, and the one we are focusing on here, is reducing the costs to science, and scientists, such that those considering to invest in software engineering are not faced with prohibitive hurdles. We suggest that creating an effective infrastructure for scientists for securing software engineering services, at the time, level, and extent that is appropriate, could remove many of those hurdles, and that a non-profit center-of-excellence model would be a cost-effective way to provide such an infrastructure.

More specifically, one of the center's main objectives would be to shift, from the scientist's perspective, investment in software engineering from a problem of justifying, identifying, recruiting, and retaining highly skilled human resources, to one that is akin to obtaining access to shared instrumentation as a service. When it comes to hardware too complex or expensive to operate for individual labs, there is ample precedence in science for the principle of benefiting from economies of scale by pooling resources in local, national, or sometimes even international core service facilities. This not only relieves investigators from the continuous evaluation, testing, operation, and maintenance of equipment technology, but also drastically broadens access to the technology to the many scientists who are not part of a well-funded big science project and who would need the technology for only a small part of its full capacity or life cycle.

The center we propose would apply the same principle to software engineering in the long tail of scientific software, the vast majority of which needs far less than a full-time engineer. The center would pool the small and otherwise disparate fractions of software engineering effort. It would recruit personnel with deep training in software engineering and knowledge across a scientific discipline sufficient to collaborate effectively with scientists across a domain (so-called T-skilled personnel). The center would provide an attractive and stimulating work environment that fosters continuous knowledge exchange among its personnel across projects, domains, and disciplines. The center would offer an alternative career track with competitive compensation that fosters talent retention and helps stem the continued loss of skilled research programmers to industry.

Scientists would obtain the services of the center by contracting outcome-oriented support packages, rather than leasing hourly staff time. Outcomes could include usability testing and resulting user-interface changes that improves the experience of new users; a code design that is more likely to be instantly understood by another developer; or implementing an algorithm in a way that removes a scalability constraint. Packages could be obtained at the time most appropriate for the targeted outcome. For example, making an algorithm more scalable may not be meaningful until the scientist has ascertained its principal usefulness for his research question. Packages can be negotiated upfront, allowing investigators to budget their costs on grant applications. This would redirect review panels from quibbling over individuals' pay rates to evaluating specifically whether the quoted software engineering outcomes are justified or not.

This model would also allow funding agencies to set more specific expectations in regard to outcomes promoting future community adoption and self-sustainability of a software tool. These could be technical in nature, such as well-designed APIs and a plugin-friendly architecture; socially oriented, such as community engagement, building, and responsiveness; or they could be about determining suitable future sources of revenue, such as identifying business models or running crowdfunding campaigns. The center would have strong expertise in each of these areas, either among its own staff, or by maintaining active liaisons to other initiatives that do. Thus, such activities may all be part of the support package scientists can contract. The package contract model also allows the center's community support commitments to extend beyond the official end of a project's grant funding, providing more time to transition a software tool to community ownership.

The center we envision would have far too little, and too highly trained resources to provide scientific programming as a service at large. Its services therefore do not substitute for increasing programming literacy and awareness of good coding practices among scientists [5]; instead, its long-term viability would depend on the success of corresponding initiatives such as the programming boot camps under the Software Carpentry brand [6]. The center would thus actively participate in Software Carpentry workshops or those of similar advocacy groups, such as the Software Sustainability Institute [7]. Likewise, the center's services would not replace but be aimed at promoting contributions from developers in the community, and hence the center's services would be contingent on a commitment to open-source and open development, using platforms such as Github that foster contribution through social coding features. In the same vein, the center could act as an umbrella mentoring organization in remote programming summer internship programs such as the Google Summer of Code, and it could sponsor additional internships on its own.

Importantly, the center would complement, assist, and train the individual scientist-developer rather than taking away from their intellectual and other ownership in the long tail tools. The center's premise is that its service would make scientists and scientist-developers more effective and more productive, by allowing them to focus on the parts for which they have unique expertise. The center's staff would collaborate closely with domain scientists and developers, including being at least periodically physically embedded with the research group. Interestingly, the informatics support teams at several US-based scientific synthesis centers already serve in some of the capacities outlined here, with considerable impact on the science sponsored by those centers [8].

Once fully running, basic operation and staff of the center would be sustained through the revenue stream from the support packages that scientists contract, which in addition to grant funding could also be sponsored through institutional subscriptions and retainer contracts. Getting the center off the ground will require a start-up grant. Aside from establishing a governance structure and a business plan, the center would need start-up funds to hire an initial 2-3 staff to lend their expertise to a few well-selected software projects that would enable the center to demonstrate its value and the potential impact of its founding principles. Projects most suitable for this phase will have wide recognition for their technical capabilities, but will have experienced comparably low community uptake, by users as well as developers, due to usability, performance, code design, architecture, and other software engineering deficiencies. Project candidates could be solicited from communities of select domains in an open call.

## Conclusion

The center we propose will by no means address all issues of software sustainability in science. However, by turning software engineering into shareable instrumentation, it would help to make the benefits of skilled engineering broadly accessible to the often small projects in the long tail, and it would enable domain scientists to easily and transparently invest in better prospects for future community uptake. Such investment may not be cheap, but with the pervasive reliance of science on computation, the continuing churn of software and of the talent that develops it have a considerable cost to science, too. Software that is sustained longer and reused more often benefits not only its creator, but accelerates science as a whole, and the people who know how to improve reuse deserve a career track.

## Acknowledgements

We thank the participants of a Birds-of-a-Feather session titled "Software Maintenance Foundation" at the 2013 iEvoBio Conference, some of the findings of which inspired this paper.